\documentclass[a4paper,11pt]{article}
\pdfoutput=1 

\usepackage{jinstpub} 

\title{\boldmath A versatile cryogenic system for liquid argon detectors}

\author[a,1]{G. Grauso,\note{Corresponding author.}}
\author[a]{A. Basco,}
\author[a]{N. Canci,}
\author[a]{R. de Asmundis,}
\author[a,b]{F. Di Capua,}
\author[a,b]{G. Matteucci,}
\author[a,b]{Y.~Suvorov,}
\author[a,b]{G. Fiorillo}

\affiliation[a]{INFN - Napoli,\\Napoli 80126, Italy}
\affiliation[b]{Physics Department, Universit\`a degli Studi di Napoli Federico II,\\Napoli 80126, Italy}

\emailAdd{gianfrancesco.grauso@na.infn.it}

\abstract{Detectors for direct dark matter search using noble gases in liquid phase as detection medium need to be coupled to liquefaction, purification and recirculation systems.
A dedicated cryogenic system has been assembled and operated at the INFN-Naples cryogenic laboratory with the aim to liquefy and purify the argon used as active target in liquid argon detectors to study the scintillation and ionization signals detected by large SiPMs arrays.\\
The cryogenic system is mainly composed of a double wall cryostat hosting the detector, a purification stage to reduce the impurities below one part per billion level, a condenser to liquefy the argon, a recirculation gas panel connected to the cryostat equipped with a custom gas pump.
The main features of the cryogenic system are reported as well as the performances, long term operations and stability in terms of the most relevant thermodynamic parameters.}


\collaboration[c]{on behalf of the DarkSide collaboration}

\proceeding{LIDINE2022\\
  September 21-23, 2022\\
  University of Warsaw Library}

\begin{document}
\maketitle
\flushbottom

\section{Introduction}
\label{sec:intro}

Noble gases (argon, xenon, neon) in liquid phase are widely used as active medium in the detectors for direct dark matter search and neutrino experiments.
Argon in particular represents a good sensitive material to this purpose. 
In fact liquid argon (LAr) technology is supported at industrial level to realize large mass detectors; argon provides high scintillation photon yield\footnote{Photon yield $\simeq$40000~$\gamma$/MeV for a minimum ionizing particle}, powerful background rejection through efficient event discrimination methods with the ionization and scintillation processes produced by particle interactions, ease of purification and high abundance ($\sim$1\% content in the atmosphere) at reasonable~cost \cite{doke1}.\\
Because of the need for thermodynamic stability and due to the presence of impurities coming from the detector materials, noble liquid apparatuses need liquefying, purifying and recirculation cryogenic systems. \cite{acciarri1, acciarri2}.
Within the framework of the DarkSide (DS) project a double phase liquid argon Time Projection Chamber (LAr TPC) detector, hosted in a double wall cryostat, has been built and installed at INFN-Naples cryogenic laboratory with the aim to study the scintillation and ionization signals detected by a large array of SiPMs \cite{dstdr, hitachi}. 
To fill, liquify and purify the argon as well as to control and keep the detector stable in terms of pressure and temperature a dedicated custom cryogenic system has been built and commissioned.

%


\section{Cryogenic system description}

\begin{figure}[h!]
\begin{center}
\includegraphics*[width=5.5cm,angle=0]{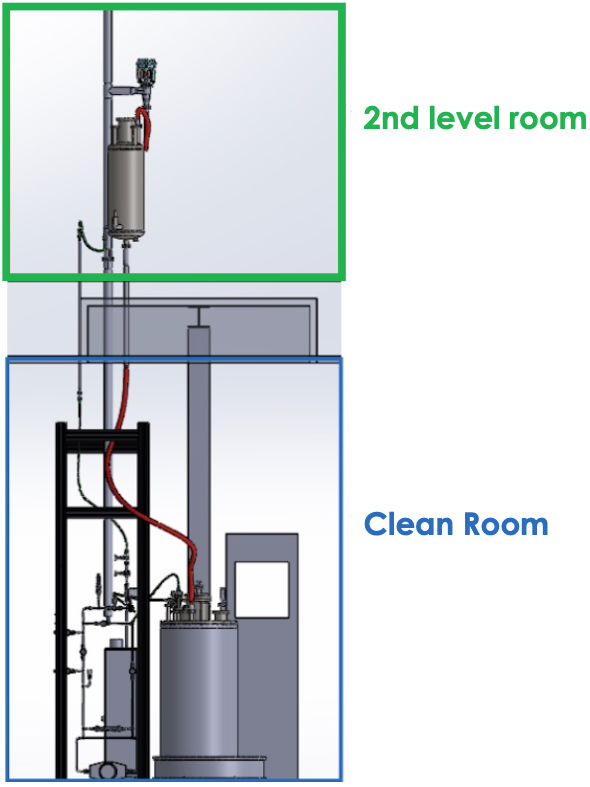}
\includegraphics*[width=9.25cm,angle=0]{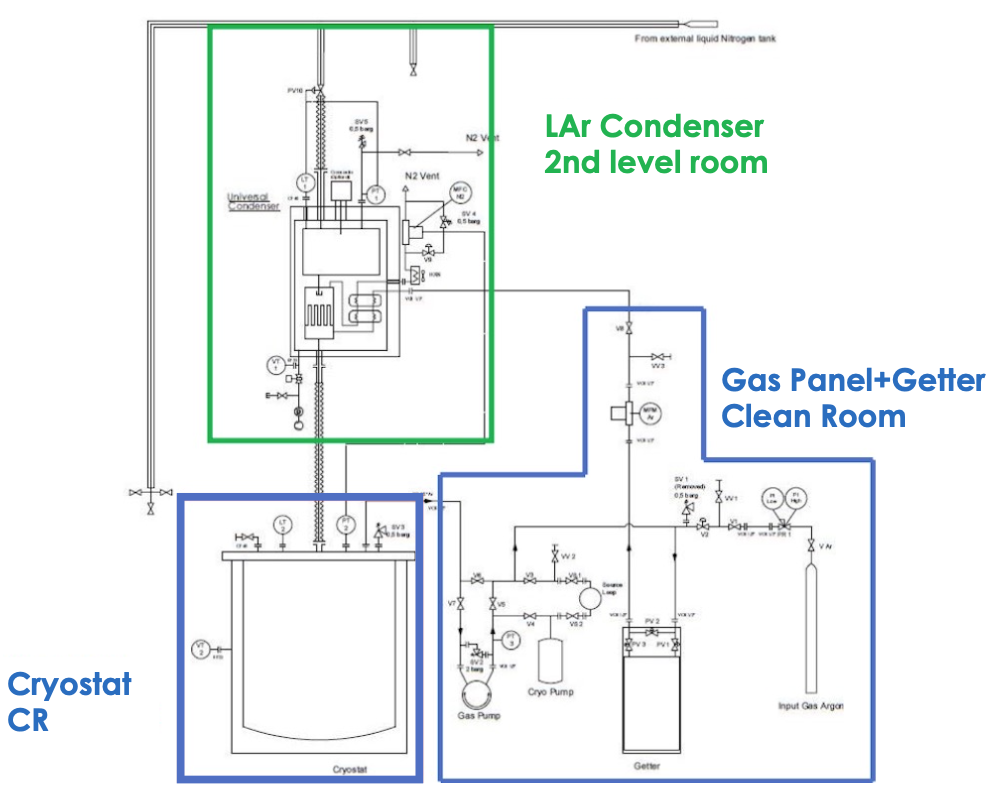}
 \caption{Model of the cryogenic system located on two floors [Left] and system P\&ID [Right].}
\label{fig:system}
\end{center}
\end{figure}

The custom cryogenic system is a combined setup made of several subsystems and can be divided in two segments located respectively inside the clean room and on the second level of the cryogenic laboratory as shown in fig.~\ref{fig:system}~[left].
The clean room hosts the double wall cryostats used for the test setup deployment, the stainless steel piping gas panel with the valves and the instrumentation for the argon flow control and monitoring, the gas recirculation pump and the getter for the argon gas purification. The condenser is placed at the second level floor above the clean room. This device, which allows to managing two separated loops for argon and nitrogen, is connected to the cryostat with a custom-made flexible double-wall transfer line and with a solid cryogenic transfer line to the external storage liquid nitrogen (LN) tank.\\

%
%
%
%
%
%
The working principle of the system is based on two splitted loops for Ar and N$_{2}$ as can be inferred by the schematics reported in the Appendix~\ref{sec:P_ID} showing the P\&ID of the system.   
By opening the manual valve (HW07) and the electric valve (EV04), the liquid nitrogen coming from the supply tank flows to a small LN buffer tank inside the condenser where the flow rate of the incoming LN is regulated by a proportional pneumatic valve (pV10) governed by a proportional integral derivative (PID) controller set on the nitrogen level. The argon liquefaction process is made of a pre-cooling phase taking place in two "brazed plate" heat exchangers, and of the effective liquefaction occurring in a "tube and shell" heat exchanger.
After the heat exchange the nitrogen in gas phase with a flow rate controlled by a mass flow controller (MFC) is flushed out into the vent line. 

At the beginning of the filling process the gaseous argon (GAr) coming from the bottle flows to the gas panel, gets directed to the getter and then to the condenser. The liquified argon drops  to the cryostat by gravity.
The liquefied argon in the cryostat evaporates due to the heat exchange with external ambient and the internal heat load produced by the experimental devices.
The evaporated argon gets pumped through the getter for purification, sent to the condenser to be liquefied again and reintegrated into the cryostat. 
The rate of GAr liquefaction inside the condenser affects the pressure of the Ar in the cryostat. 
The cryostat pressure is kept stable by an active PID control capable of regulating the gaseous nitrogen output flow, adjusting the  cooling power and determining the condenser efficiency.



The condenser has been designed to operate in two different configurations. In the first configuration, a continuous direct supply of liquid nitrogen is used as heat exchanger fluid for the argon, while in the second one a cold head cryocooler can be installed on the dedicated top flange of the condenser to liquefy the gaseous nitrogen and reintegrate the amount of LN spent during the argon liquefaction process. The latter configuration would reduce the nitrogen consumption and would improve the efficiency of the system.

%
The getter used in the system is the PS$4$-MT$50$-R SAES MonoTorr rare-gas hot purifier specifically designed to provide ultra-high purity gas for semiconductor applications \cite{saes}. The patented getter alloy operated at elevated temperatures (up to $300^\circ$C) removes impurities by forming irreversible chemical bonds reducing the outlet impurity levels for O$_2$, H$_2$O, CO, CO$_2$, H$_2$, N$_2$ and CH$_4$ to < 1 ppb level as required for ionization detection~\cite{acciarri2, bakale, icarus, buckley}.




%

The cryogenic system can be controlled remotely during operations in a mostly automatic way.
In fact, once the cryostat is filled and the recirculation started the active components of the cryogenic system are automatically controlled on the basis of the pressure in the LAr cryostat and of the LN level in the condenser buffer tank by a Slow Control (SC) system. 
The SC is based on a National Instrument LabVIEW Real-Time Application software running on a CompactRIO controller that collects data from the instrumentation (pressure trensducers, level transducers, mass flow meter and mass flow controller)~\cite{ni}.



\section{Cryogenic system commissioning phase}

\begin{figure}[h!]
\begin{center}
\includegraphics*[width=15.5cm,angle=0]{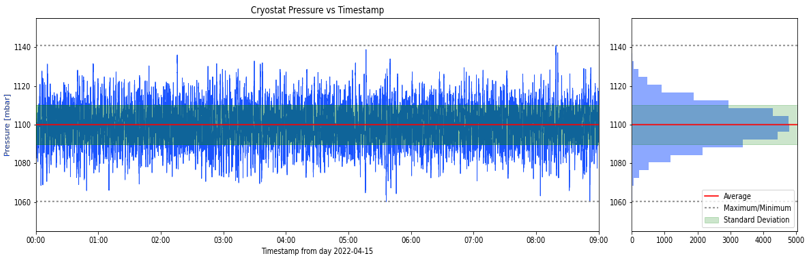}
 \caption{Cryostat pressure stability trend during the first commissioning test phase.}
\label{fig:test_1}
\end{center}
\end{figure}

In the first commissioning test a small 12.9~l double wall vacuum insulated dewar has been used in order to reduce the volume and filling time, allowing for a fast feedback on the cryogenic system features. 
A Metal Bellows gas pump (Senior Aerospace MB-602) was installed for the argon recirculation. Since the pump motor speed is fixed (2875~rpm) a manual valve has been placed at the pump outlet to regulate the flow rate.
The cryostat was filled by liquifying gaseous Ar~6.0 (1~ppm impurities in total) from pressurized bottles.
After LAr filling the recirculation stage was started including the getter in the loop for argon purification.
This commissioning test was mainly focused on the calibration and adjustment of the several active components of the system in order to operate with a stable pressure in the cryostat.
With this configuration the minimum pressure fluctuation around the set point of 1100~mbar was found to be lower than about 40~mbar with a standard deviation of 10~mbar, as can be seen in fig.~\ref{fig:test_1}.

%
%
%
%
%
%

\section{System upgrade and comparative test}

\begin{figure}[h!]
\begin{center}
\includegraphics*[width=15.5cm,angle=0]{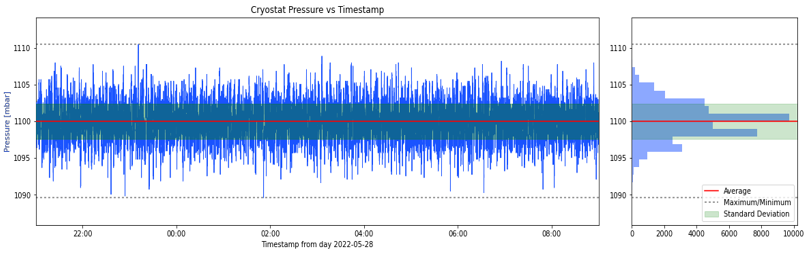}
 \caption{Behavior of the cryostat pressure stability after the changes in the cryogenic system.}
\label{fig:test_2}
\end{center}
\end{figure}

\begin{figure}[h!]
\begin{center}
\includegraphics*[width=15.5cm,angle=0]{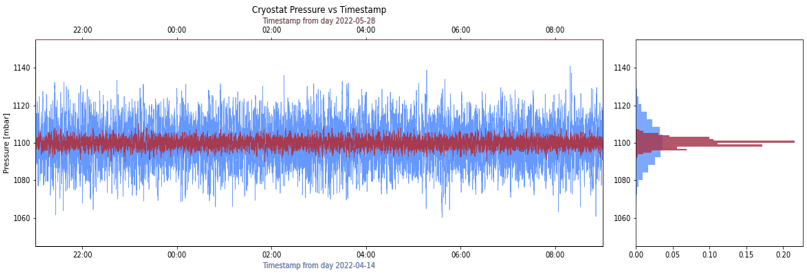}
 \caption{Comparison of the pressure trends as function of the time during the performed tests. (Blue: First commissionig test phase, Red: Second commissioning test phase}
\label{fig:test_comparison}
\end{center}
\end{figure}

After the fine tuning of the control and operational parameters a new test has been performed with the aim to improve the performance of the system, mainly in terms of pressure variation after some changes in the system configuration.
For this purpose a larger 300~l cryostat has been used for the test.
The previously used MB 602 gas pump was changed with a tunable speed custom QDrive recirculation pump. Two expansion tanks were added to reduce the mechanical vibration and the possibile gas pulsation produced during the recirculation phase.
In addition, to speed up the process the cryostat has been directly filled with LAr~5.0 purified with an in-line cartridge specifically designed for liquid argon. To avoid the risk of contamination, the getter was inactive during the test and excluded from the argon loop. 
During the whole test period the system ran at stable conditions obtaining a maximum pressure variation around the set point lower than 10~mbar, with a standard deviation of 2.4~mbar as shown in fig.~\ref{fig:test_2}.
As a comparison, the behaviour of the pressure as a function of time have been reported in fig.~\ref{fig:test_comparison} and tab.~\ref{tab}, clearly demonstrating the reduction of the pressure variation in a range of $\pm$10~mbar around the set point of $\simeq$1100~mbar after the changes in the cryogenic system.
%
%
%
%
%
%
%
%
%
%
%
%
%

\begin{table}[h]
\centering
\begin{tabular}{lcccc}
\hline
\hline
& Average [mbar] & Std. Deviation [mbar]& Max [mbar] & Min [mbar] \\
\hline
\hline
First test phase & 1100 & 10.28 & 1141 & 1060\\
Second test phase & 1100 & 2.41 & 1110 & 1090\\
\hline
\hline

\end{tabular}
\caption{\label{tab} Comparison of the pressure trends during the two test phases}
\end{table}

\newpage

\section{Conclusions}

A versatile custom-made cryogenic system has been installed and commissioned at INFN Naples cryogenic laboratory aiming to fill in a fast, efficient and stable way, recirculate and purify the experimental set up with gaseous argon, and able to keep the cryostat conditions stable over the long period of data-taking. 
After the commissioning phase needed for some fine tuning of the operational parameters and devices, an important improvement of the system was performed that lead to a sensible reduction of pressure variations in the cryostat of the order of $\pm$10~mbar over a pressure reference working point of 1100~mbar. The pressure stability shown during the test allows to keep the cryogenic parameters of the liquid argon detector stable, and to explore more accurately the main detector features related to the scintillation and ionization processes.

\newpage

\appendix
\section{Cryogenic System P\&ID}
\label{sec:P_ID}
\begin{figure}[h!]
\begin{center}
\includegraphics*[scale=0.76,angle=90,trim=70 10 45 10,clip]{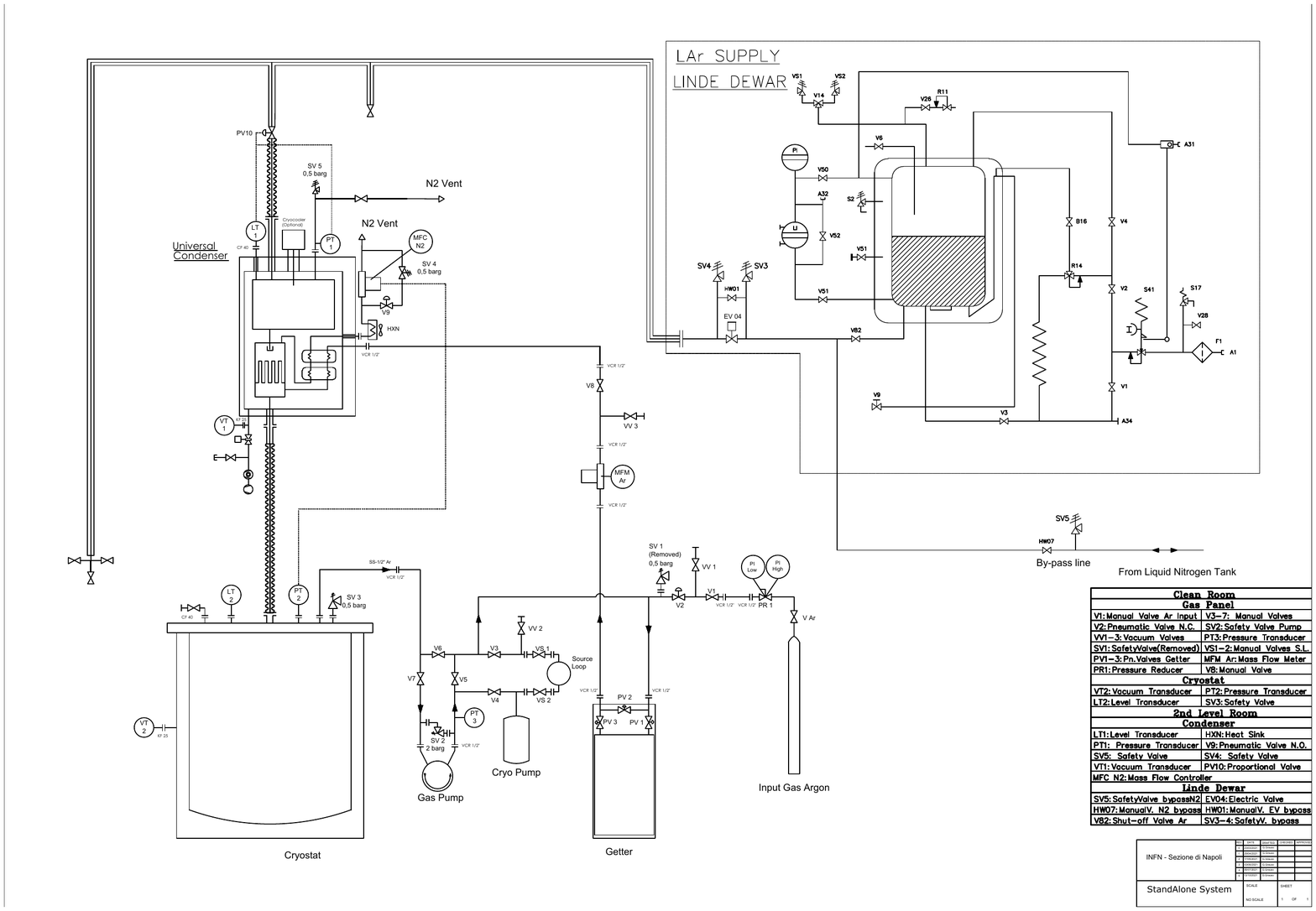}

\label{fig:P_ID}
\end{center}
\end{figure}

\acknowledgments


The authors would like to thank Dr.~Hanguo~Wang for the concept design of the condenser and the fruitful discussions, Eng.~Cary~Kendziora, Dr.~Alexander~Kish, Prof.~Yi~Wang, Dr.~Xiang Xiao, Dr.~Tom~Thorpe for the productive suggestions and indication on the cryogenic system, Eng.~Marco~Carlini for the constant help on the engineering design and Gennaro~Tortone for the monitoring system implementation.



\end{document}